\RequirePackage{fix-cm}
\documentclass[twocolumn]{svjour3}          
\smartqed  
\usepackage{graphicx}
 \usepackage{mathptmx}      
\usepackage{amsmath}
\usepackage{amssymb}
\usepackage{lineno}
\usepackage{natbib}
\usepackage{color}
\begin{document}

\title{Predicting regional and pan-Arctic sea ice anomalies with kernel analog forecasting
}


\author{Darin Comeau  \and Dimitrios Giannakis \and Zhizhen Zhao \and Andrew J. Majda
}


\institute{D. Comeau \at
 Center for Atmosphere Ocean Science, Courant Institute of Mathematical Sciences, New York University, New York, NY.\\
              \email{dcomeau@lanl.gov}           
             \emph{Present address:} of D. Comeau:\\ 
              Climate, Ocean, and Sea Ice Modeling Group, Computational Physics and Methods Group (CCS-2), Los Alamos National Laboratory, Los Alamos, NM.
           \and D. Giannakis \at
Center for Atmosphere Ocean Science, Courant Institute of Mathematical Sciences, New York University, New York, NY.
	\and Z. Zhao \at 
Department of Electrical and Computer Engineering, Coordinated Science Laboratory, University of Illinois at Urbana-Champaign, Urbana, IL.
           \and A. Majda \at
Center for Atmosphere Ocean Science, Courant Institute of Mathematical Sciences, New York University, New York, NY.              
}

\date{Received: date / Accepted: date}

\maketitle

\begin{abstract}
Predicting Arctic sea ice extent is a notoriously difficult forecasting problem, even for lead times as short as one month. Motivated by Arctic intraannual variability phenomena such as reemergence of sea surface temperature and sea ice anomalies, we use a prediction approach for sea ice anomalies based on analog forecasting. Traditional analog forecasting relies on identifying a single analog in a historical record, usually by minimizing Euclidean distance, and forming a forecast from the analog's historical trajectory. Here an ensemble of analogs is used to make forecasts, where the ensemble weights are determined by a dynamics-adapted similarity kernel, which takes into account the nonlinear geometry on the underlying data manifold. We apply this method for forecasting pan-Arctic and regional sea ice area and volume anomalies from multi-century climate model data, and in many cases find improvement over the benchmark damped persistence forecast. Examples of success include the 3--6 month lead time prediction of Arctic sea ice area, the winter sea ice area prediction of some marginal ice zone seas, and the 3--12 month lead time prediction of sea ice volume anomalies in many central Arctic basins. We discuss possible connections between KAF success and sea ice reemergence, and find KAF to be successful in regions and seasons exhibiting high interannual variability.
\end{abstract}


\section{Introduction}
\label{sec:intro}
\par Predicting the climate state of the Arctic, particularly with regards to sea ice extent, has been a subject of increased recent interest, in part driven by record-breaking minima in September sea ice extent in 2007 and again in 2012. As new areas of the Arctic become accessible, this has increasingly become an important practical problem in addition to a scientific one, e.g., for navigating shipping routes \citep{smith2013new}. Many different approaches have been recently employed to address Arctic sea ice prediction, including both statistical frameworks \citep{lindsay2008seasonal, wang2016predicting}, and dynamical models \citep{blanchard2011persistence, blanchard2011influence, chevallier2013seasonal, tietsche2013predictability, tietsche2014seasonal, day2014pan, zhang2008ensemble,sigmond2013seasonal, wang2013seasonal}. These methods have varying degrees of success in predicting sea ice area or extent (area with at least 15\% sea ice concentration) and sea ice volume for the pan-Arctic or regions of interest. Indeed, in sea ice prediction, current generation numerical models and data assimilation systems have little additional skill beyond simple persistence or damped persistence forecasts \citep{blanchard2015model}.

\par Following the 2007 September sea ice extent minimum, the Study of Environmental Arctic Change (SEARCH) began soliciting forecasts of September sea ice extent for the Sea Ice Outlook (SIO), which since 2013 has been handled by the Sea Ice Prediction Network (SIPN). They have found that year to year variability, rather than methods, dominate the ensemble's success, and that extreme years are in general less predictable \citep{stroeve2014predicting}. The forecasts, given at one to three lead month times, had particular difficulty with the record low extent of September 2012 and the subsequent September 2013, which saw a partial recovery in sea ice extent. A more recent study of SIO model forecasts by \citet{blanchard2015model} highlighted the importance of model uncertainty on predictability by performing an initial condition perturbation experiment and finding wide spread among models' response.

\par Accurately predicting aspects of Arctic sea ice is made difficult by a number of factors, notably that the mean state of the Arctic is changing \citep{stroeve2014predicting,guemas2016review}. In particular, statistical prediction methods based on historical relationships have difficulty in predicting sea ice in a changing Arctic mean state \citep{holland2011changing, stroeve2014predicting}. As ice becomes thinner, previous studies have shown it becomes more variable in extent \citep{holland2011inherent,goosse2009increased,holland2008role}, and is hypothesized to have lower predictability \citep{blanchard2011persistence, holland2008role, holland2011inherent, germe2014interannual}. Since the satellite record began in 1979, every month has exhibited a year to year downward trend in sea ice extent, the largest being for September \citep{stroeve2012trends}. Moreover, as thicker multiyear ice is replaced by thinner, younger ice, the trends steepen \citep{stroeve2012trends}. Ice thickness data is seen as offering key predictive information for both sea ice area and extent \citep{bushuk2017summer,blanchard2014characteristics, chevallier2012role, lindsay2008seasonal, wang2013seasonal}, but such observational data sets do not yet exist in uniform spatial and temporal coverage. However some observations are available from various satellites such as ICESat \citep{kwok2007ice} (and the upcoming ICESat-2), CryoSat-2 \citep{tilling2016near}, and SMOS \citep{kaleschke2012sea}. There is also the commonly used assimilation product based on the Pan-Arctic Ice Ocean Modeling and Assimilation System (PIOMAS) \citep{zhang2003modeling}, which produces sea ice volume data by assimilating observations of sea ice concentration with a regional ice-ocean model. Ice age, in particular area of ice of a certain age, is also seen as an important predictor, also of which there is no reliable observational record \citep{stroeve2012trends}. 

\par Both dynamic and thermodynamic elements factor into sea ice predictability. Locally, ice thickness predictability in the Arctic is dominated by dynamic, rather than thermodynamic properties \citep{blanchard2014characteristics, tietsche2014seasonal}. On the other hand, limits on September sea ice extent are primarily thermodynamic (related to the amount of open water formation in melt season), whereas dynamic induced anomalies have a  smaller influence, except in a thin ice regime \citep{holland2011inherent}. Improvement in melt-pond parameterizations in the sea ice model Community Ice CodE (CICE) \citep{holland2012improved} has yielded skill in predicting September sea ice extent \citep{schroder2014september}, demonstrating potential predictive yield in improving process models.

\par Chaotic atmosphere variability also places an inherent, and perhaps dominant, limit on sea ice predictability, both through dynamic effects via redistribution of sea ice \citep{day2014pan, holland2011inherent, blanchard2011influence, ogi2010influence}, and thermodynamic effects on ocean conditions \citep{tietsche2016atmospheric}. While correlations between the Arctic Oscillation (AO) and summer ice extent historically have been high \citep{rigor2002response,rigor2004variations}, these correlations have become weaker in recent years. The sign of the AO has changed while sea ice extent has continued to decline, suggesting this climate index may not be a very predictive atmospheric metric for sea ice \citep{holland2011changing,ogi2010influence}. Nevertheless, summer atmospheric conditions have a strong impact on sea ice extent, particularly for thinning sea ice, and may contain more predictive skill than sea ice thickness in terms of predicting the September ice extent minimum \citep{stroeve2014predicting}. The ocean temperature at depth has also been found to be an important predictor for sea ice extent \citep{lindsay2008seasonal} through storage of subsurface heat anomalies.

\par Another aspect of sea ice variability with a seasonal dependence is sea ice anomaly reemergence, a phenomenon where anomalies at one time reappear several months later, made evident by high lagged correlations following low correlations at shorter time lags. This behavior has been found in both models and observations \citep{blanchard2011persistence}. Reemergence phenomena fall into two categories. One type is where anomalies from a melt season reemerge in the subsequent growth season, typically found in marginal ice zones, and is governed by ocean and large-scale atmospheric conditions \citep{blanchard2011persistence,bushuk2014reemergence,bushuk2015sea,bushuk2015arctic,bushuk2017seasonality}. Another type is where anomalies from a growth season reemerge in the subsequent melt season, typically found in central Arctic regions that exhibit perennial sea ice, and is driven by sea ice thickness \citep{blanchard2011persistence,day2014pan,bushuk2017seasonality}. These observed phenomena may provide a promising source of sea ice predictability, which \citet{day2014pan} found to be robust in several GCMs. 

\par The problem of sea ice prediction becomes both of more practical use, while becoming more difficult, as we move from the pan-Arctic to regional scale, where local ice advection across regional boundaries and small scale influences on sea ice processes become important \citep{blanchard2014characteristics}. Seasonal ice zones have different factors that control predictability, including reemergence in the Pacific marginal ice zones, and the regulating effect of the North Atlantic on the Atlantic marginal ice zones \citep{yeager2015predicted, koenigk2009seasonal}. For sea ice thickness, persistence in the central Arctic basins leads to higher predictability than in other seasonal ice regions \citep{day2014pan}. In addition to the September sea ice extent metric, there has been increased focus on predicting regional sea ice advance and retreat dates (e.g. \citet{sigmond2016skillful}), which are now included as part of the SIO solicitation, as well as predicting the Arctic sea ice edge \citep{goessling2016predictability}. Seasonality also plays a strong role in predictability, e.g. SST conditions in the North Atlantic can lead to high predictability of winter sea ice extent \citep{yeager2015predicted}, whereas the summer melt season provides a barrier to predictability, limiting the skill of forecasts initialized in late spring \citep{day2014pan}.

\par The timescales of Arctic sea ice predictability vary across studies, depending on the measure of forecast skill and the initial month of prediction (among other factors), but generally fall in the 3--6 month range \citep{blanchard2015model,guemas2016review,tietsche2014seasonal,chevallier2012role}, with potential predictability extending to 2--3 years \citep{tietsche2014seasonal,blanchard2011influence}. While \citet{lindsay2008seasonal} found that most predictive information in the ice-ocean system is lost for lead times greater than 3 months, \citet{blanchard2011persistence} found pan-Arctic sea ice area predictable in a perfect model framework for 1--2 years, and sea ice volume up to 3--4 years. It has been found that predicting the state of sea ice in the spring is particularly difficult, with most of the predictive skill coming from fall persistence \citep{wang2013seasonal, holland2011inherent}, and that for detrended data, March sea ice extent is largely uncorrelated with the following September sea ice extent \citep{blanchard2011persistence, stroeve2014predicting}.

\par While it is common to use computationally expensive dynamical models for forecasting methods, in recent years the SIO has received statistical forecasts in roughly equal number to those based on dynamic ocean-ice models \citep{hamilton2016400}, and there is promise in utilizing nonparametric statistical techniques. Among these, analog forecasting is an idea dating back to \citet{lorenz1969atmospheric}, where a prediction is made by identifying an appropriate historical analog to a given initial state, and using the analog's trajectory in the historical record to make a forecast of the present state. While this is attractive as a fully non-parametric, data-driven approach, a drawback of traditional analog forecasting is that it relies upon a single analog, usually identified by Euclidean distance, possibly introducing highly discontinuous behavior into the forecasting scheme. This can be improved upon by selecting an ensemble of analogs, and taking a weighted average of the associated trajectories. Another potential drawback is that a large number of historical data may be needed in order to identify appropriate analogs, particularly if the number of degrees of freedom is high \citep{van1994searching}, as is often the case in climate applications. Nevertheless, analog forecasting in some form has been used in numerous climate applications \citep{drosdowsky1994analog, xavier2007analog, alessandrini2015novel}, including in an ensemble approach \citep{atencia2015comparison, liu2017improving}. Given there are sources of sea ice predictability from the ocean, atmosphere, and sea ice itself \citep{guemas2016review}, a data-driven approach such as analog forecasting may be able to exploit complex coupled-system dynamics encoded in GCM data and provide skill in such a prediction problem.

\par In \citet{zhao2016analog}, this idea was extended upon by assigning ensemble weights derived from a dynamics-adapted kernel, constructed in such a way as to give preferential weight to states with similar dynamics, referred to as kernel analog forecasting (KAF). Modes of variability intrinsic to the data analysis, as eigenfunctions of the kernel operator, are extracted with clean timescale separation and inherent predictability through the encoding of dynamic information in the analysis. KAF has been used in forecasting modes representing the Pacific Decadal Oscillation (PDO) and North Pacific Gyre Oscillation (NPGO) \citep{zhao2016analog}, in which cases it was shown to be more skillful in potential predictability than parametric regression forecasting methods \citep{comeau2017data}. More recently, KAF has been used in forecasting variability in the tropics attributed to the Madden-Julian oscillation and boreal summer intraseasonal oscillation using observations \citep{alexander2017kernel}.

\par These examples demonstrate KAF exhibiting skill in predicting modes intrinsic to the data analysis, that is to say eigenfunctions of the kernel operator (similar to EOF principal components). A more practical problem is in forecasting observables that do not rely on a particular data analysis method, but rather can be physically observed, e.g. Arctic sea ice extent anomalies \citep{comeau2017data}. The aim of this study is to extend upon \citet{comeau2017data} and assess the skill of KAF in predicting Arctic sea ice anomalies on various spatial and temporal scales in order to identify where and when we may (or may not) have predictive skill. Since, as with every statistical technique, the utility of KAF depends upon the availability of an appropriately rich training record, we examine predictability in a long control run of a coupled GCM to establish a baseline of KAF predictive skill in predicting the internal dynamics attributed to natural variability. This allows us to estimate the upper limits of skill with this method in a statistically robust manner. We consider various combinations of predictor variables including sea ice concentration (SIC), sea surface temperature (SST), sea ice thickness (SIT), and sea level pressure (SLP) data to assess which variables contain the most useful predictive information. The important consideration of statistical prediction in the presence of a changing climate is beyond the scope of this work, though in Sect.~\ref{sec:conclusions} we make suggestions of how to address this issue. 

\par The rest of this paper is structured as follows: The KAF method is described in Sect.~\ref{sec:methods}. The data and experimental setup are described in Sect.~\ref{sec:data}, with the associated results in Sect.~\ref{sec:results}. Discussion and concluding remarks are given in Sect.~\ref{sec:conclusions}.

\section{Methods}
\label{sec:methods}
\par The KAF method \citep{zhao2016analog,comeau2017data,alexander2017kernel}, is designed to address the difficult task of prediction using massive data sets sampled from a complex nonlinear dynamical system in a very large state space. The motivating idea is to encode information from the underlying dynamics of the system into a kernel function, which is an exponentially decaying pairwise measure of similarity that plays an analogous role to the covariance operator in principal components analysis. At the outset, during the training phase we have access to an $n$ sample size time-ordered training data set $\{x_1,\ldots,x_n\}$ and the corresponding values $\{f_1,\ldots,f_n\}$ of a prediction observable. In our applications, the target observable is the aggregate sea ice anomaly in extent or volume over some region, and the training data are monthly averaged gridded climate variables. The main steps in KAF, outlined in detail below, are 1) perform Takens embedding of the data, 2) evaluate a dynamics-adapted similarity kernel on the embedded data, and 3) use weights from this kernel to make a forecast of an observable via out-of-sample extension formed by a weighted iterated sum.

\subsection{Takens embedding}
	\par The first step in our analysis is to construct a new state variable through time-lagged embedding. For an embedding window of length $q$, which will depend on the time scale of our observable of interest, and a spatiotemporal series $z_1, z_2,\ldots z_n$ with $z_i\in\mathbb{R}^d$, where $d$ is the number of spatial (grid) points and $i$ is the time index, we form a data set of $x_i$ in lagged-embedded space (also called Takens embedding space) by
	
	\begin{equation*}
	x_i = \left(z_i, z_{i-1}, \ldots, z_{i-(q-1)}\right)\in\mathbb{R}^{qd}.
	\end{equation*}	
The utility of this embedding is that it recovers the topology of the attractor of the underlying dynamical system through partial observations (the $z_i$'s) \citep{packard1980geometry, takens1981detecting, broomhead1986extracting, sauer1991embedology, robinson2005topological, deyle2011generalized}. The choice of the embedding window $q$ should be chosen long enough to capture the time-scales of interest, but not so long as to reduce the discriminating power of the kernel in determining locality. 

\subsection{Dynamics-adapted kernels}
\label{subsec:kernels}
	\par The collection of data points $\{x_1,\ldots,x_n\}$ can be thought of as lying on a manifold nonlinearly embedded in data space $\mathbb{R}^{qd}$. We will endow this manifold with a geometry (i.e., a means of measuring distances and angles) through a kernel function of data similarity. The kernel function we use for that purpose is from the Nonlinear Laplacian Spectral Analysis (NLSA) algorithm \citep{giannakis2012comparing, giannakis2012nonlinear, giannakis2013nonlinear, giannakis2014data}. The kernel incorporates additional dynamic information by using phase velocities $\xi_i = \|x_i - x_{i-1}\|$, thus giving higher weight to regions of data space where the data is changing rapidly (see \citet{giannakis2015dynamics} for a geometrical interpretation), and is given by
	
		\begin{equation}
		k\left(x_i,x_j\right) = \exp\left(-\frac{\|x_i - x_j\|^2}{\epsilon\xi_i\xi_j}\right).
		\label{eq:nlsa_kernel}
		\end{equation}		
In the above, $\|\cdot\|$ is the standard Euclidean norm and $\epsilon$ is a positive scale parameter. We can generalize this definition to include multiple variables $x_i = \left(x^{(1)}_i, x^{(2)}_i\right)$ \citep{bushuk2014reemergence}, possibly of different physical units, embedding windows, or grid points. For instance, the analogous kernel to \eqref{eq:nlsa_kernel} for two variables is
		
		\begin{equation}
		k\left(x_i,x_j\right) = \exp\left(-\frac{\|x^{(1)}_i - x^{(1)}_j\|^2}{\epsilon^{(1)}\xi^{(1)}_i\xi^{(1)}_j} - \frac{\|x^{(2)}_i - x^{(2)}_j\|^2}{\epsilon^{(2)}\xi^{(2)}_i\xi^{(2)}_j}\right),
		\label{eq:nlsakernel_multivariate}
		\end{equation}
and this approach can be extended to more than two variables in a similar manner. While in principle different scale parameters $\epsilon$ may be used for different variables to assign relative weights between variables within the kernel, in this analysis we use the same scale parameter for all variables. The exponential decay of the kernel in Eq.~\eqref{eq:nlsakernel_multivariate} means that very dissimilar states will carry negligible weight in our construction of historical analogs. In practice we enhance this localizing behavior further by setting small entries of $k$ to zero, thereby reducing the ensemble size. Finally, we next form row-normalized kernels,
		
	\begin{equation}
		P(x_i,x_j) = \frac{k(x_i,x_j)}{\sum_{l=1}^nk(x_i,x_l)},
		\label{eq:rowsum}
	\end{equation}
which forms a row-stochastic matrix $P$ that allows us to interpret each row as an empirical probability distribution of the second argument that depends on the data point in the first argument. 
		
\subsection{Out-of-sample extension via Laplacian pyramids}
\label{subsec:LP}
	\par Our approach of assigning a value for a function $f$ defined on a training data set $X$ to a new test value $y\notin X$ will be through an out-of-sample extension technique known as Laplacian pyramids \citep{rabin2012heterogeneous}. In our context, the training data will be a spatiotemporal data set comprised of (lagged-embedded) state vectors $x_i$ of gridded state variables (e.g. some combination of SIC, SST, SLP, and SIT), $f_i = f(x_i)$ is the function that gives us the sea ice area anomaly of the state $x_i$, and $y$ will be a new state vector (in lagged-embedded space), from which we would like to make a forecast of future sea ice area anomalies.
	\par We define a family of kernels $P_l$ by modifying the NLSA kernels $k$ in Eq.~\eqref{eq:rowsum} to have a scale parameter of the form $\sigma_0/2^l$ rather than $\epsilon$, which we denote $k_l$, and then $P_l$ is the row-sum normalized $k_l$, as in Eq.~\eqref{eq:rowsum}. This forms a multiscale family of kernels with increasing dyadic resolution in $l$ with a shape parameter $\sigma_0$.
	A function $f$ is approximated in a multiscale manner as an iterated weighted sum by $f\approx s_0 + s_1 + s_2 + \cdots$, where the first level $s_0$ and difference $d_1$ is defined by

	\begin{equation*}
	s_0(x_k) = \sum_{i=1}^nP_0(x_k,x_i)f(x_i), \qquad d_1(x_i) = f(x_i) - s_0(x_i),
	\end{equation*}
and the $l$th level decomposition $s_l$ is defined iteratively:

	\begin{equation*}
	s_l(x_k) = \sum_{i=1}^n P_l(x_k,x_i)d_l(x_i), \qquad d_l(x_i) = f(x_i) - \sum_{i=0}^{l-1}s_i(x_i).
	\end{equation*}
	
	\par Note that the sum over $i$ to obtain $s_l(x_k)$ is over the $n$ training data points. For the choice of kernels $k_l$, increasing $l$ can lead to overfitting, which we mitigate by zeroing out the diagonals of the kernels \citep{fernandez2013auto}. We set the truncation level for the iterations at level $L$ once the approximation error begins to increase in $l$. 
	Given a new data point $y$, we extend $f$ by
	
	\begin{equation*}
	\bar{s}_0(y) = \sum_{i=1}^n P_0(y,x_i)f(x_i),\qquad \bar{s}_l(y) = \sum_{i=1}^n P_l(y,x_i)d_l(x_i),
	\end{equation*}
for $L\geq1$, and assign $f$ the value
	
	\begin{equation}
	\bar{f}(y) = \sum_{l=0}^L \bar{s}_l(y).
	\label{eq:ose_lp}
	\end{equation}
	
	\par That is, we use the kernels $P_l$ with $l\leq L$ to evaluate the similarity of $y$ to points $x_i$ in the training data, and use this measure of similarity to form a weighted average of $f(x_i)$ values to define $\bar{f}(y)$. Note that there will be some reconstruction error between the out-of-sample extension value $\bar{f}(y)$ and the ground truth $f(y)$, which in general is not known.

\subsection{Kernel Analog Forecast (KAF)}
	\par Recall that in traditional analog forecasting, a forecast is made by identifying a single historical analog that is most similar to the given initial state, and using the historical analog's trajectory as the forecast.  As mentioned in Sect.~\ref{subsec:kernels}, it is convenient to think of rows of normalized kernels as empirical probability distributions in the second argument, $p_y(x) = P(y,x)$. In this setting, we can then consider taking weighted sums as taking an expectation ($\mathbb{E}$) over a probability distribution. As an example, the traditional analog forecast $\bar{f}_{TAF}$ for a lead time $\tau$ can be written as 
	
	\begin{equation*}
	\bar{f}_{TAF}(y,\tau)= \mathbb{E}_{p_y}S_\tau f = \sum_{i=1}^n p_y(x_i)f\left(x_{i+\tau}\right) = f\left(x_{j + \tau}\right),
	\end{equation*}
where $p_y = \delta_{iy}$ is the Kronecker delta distribution and $S_\tau f(x_i) = f(x_{i+\tau})$ is the shift operator. Note that $S_\tau f(x_i)$ can be evaluated since we know the time-shifted value of $f$ exactly over the training data set.	
	\par Given a new state $y$, we define our prediction $\bar{f}(y,\tau)$ for lead time $\tau$, via Laplacian pyramids, by
	
	\begin{equation}
	\bar{f}(y,\tau) = \mathbb{E}_{p_{y,0}}S_\tau f + \sum_{l=1}^L\mathbb{E}_{p_{y,l}}S_\tau d_l,
	\label{eq:kaf}
	\end{equation}
where $p_{y,l}(x) = P_l(y,x)$ corresponds to the probability distribution from the kernel at scale $l$. Note that when $\tau=0$, Eq.~\eqref{eq:kaf} reduces to the Laplacian pyramid out-of-sample extension expression for $\bar{f}(y)$ in Eq.~\eqref{eq:ose_lp}.
	\par The reconstruction error from the out-of-sample extension manifests itself in the fidelity of the forecasts as the error at time lag 0. While in our applications, knowing the full climate state $y$ allows us to compute the observable $f(y)$ exactly at time lag 0, we need the out-of-sample extension to compute the predicted observable $\bar{f}(y,\tau)$ at any time lag $\tau>0$. Hence we must contend with the initial reconstruction error, which is the difference between $f(y)$ and $\bar{f}(y,0)$. This will impact forecasts for all time lags.

\subsection{Error Metrics}
\label{subsec:error_metrics}
	\par For the purposes of defining the error metrics for predictions, let $F(y_j,\tau)$ be a general prediction of an observable $f$ of state $y_j$ at lead time $\tau$, with $f(y_{j+\tau})$ being the true value. We evaluate the performance of predictions with two aggregate error metrics, the root-mean-square error (RMSE) and pattern correlation (PC), defined as
	
	\begin{eqnarray*}
	\text{RMSE}(\tau) & = & \sqrt{\frac{1}{n'}\sum_{j=1}^{n'} \left(F(y_j,\tau) - f(y_{j+\tau})\right)^2},\\	
	\text{PC}(\tau) & = & \frac{1}{n'}\sum_{j=1}^{n'}\frac{\left(F(y_j,\tau) - \tilde{F}(y,\tau)\right)\left(f(y_{j+\tau}) - \tilde{f}(y, \tau)\right)}{\sigma_{\tilde{F}(y,\tau)}\sigma_{\tilde{f}(y,\tau)}},
	\end{eqnarray*}
	
	where 
	
	\begin{equation*}
	\tilde{F}(y,\tau) = \frac{1}{n'}\sum_{j=1}^{n'}F(y_j,\tau), \qquad \tilde{f}(y,\tau) = \frac{1}{n'}\sum_{j=1}^{n'} f(y_{j+\tau}),
	\end{equation*} 
	\begin{eqnarray*}
	\sigma^2_{\tilde{F}(y,\tau)} & = & \frac{1}{n'}\sum_{j=1}^{n'}(F(y_j,\tau) - \tilde{F}(y,\tau))^2, \\
	\sigma^2_{\tilde{f}(y,\tau)} & = & \frac{1}{n'}\sum_{j=1}^{n'}(f(y_{j+\tau}) - \tilde{f}(y,\tau))^2,
	\end{eqnarray*}
	
where averaging is over predictions formed from using testing data of length $n'$ (second portion of the data set) as initial conditions. KAF error metrics are evaluated with the predictions $F(y,\tau) = \bar{f}(y,\tau)$ (as defined in Eq.~\eqref{eq:kaf}). We use error metrics for the damped persistence forecast $F(y_j,\tau) = \beta^\tau f(y_j)$, where $\beta$ is the lag-1 autocorrelation coefficient of $f$, as our benchmark, and use a threshold of 0.5 in pattern correlation score, below which predictive skill is considered low \citep{germe2014interannual}. Given our interest in high ($>0.5$) pattern correlation and the large number of samples in our test data set $(\approx4500)$, the correlations considered are statistically significant. RMSE scores are normalized by the standard deviation of the truth (NRMSE) in the figures that follow, with NRMSE values approaching 1 indicating a loss of predictive skill.

\section{Datasets}
\label{sec:data}

\par
\begin{figure}
\centering
\noindent\includegraphics[width=84mm]{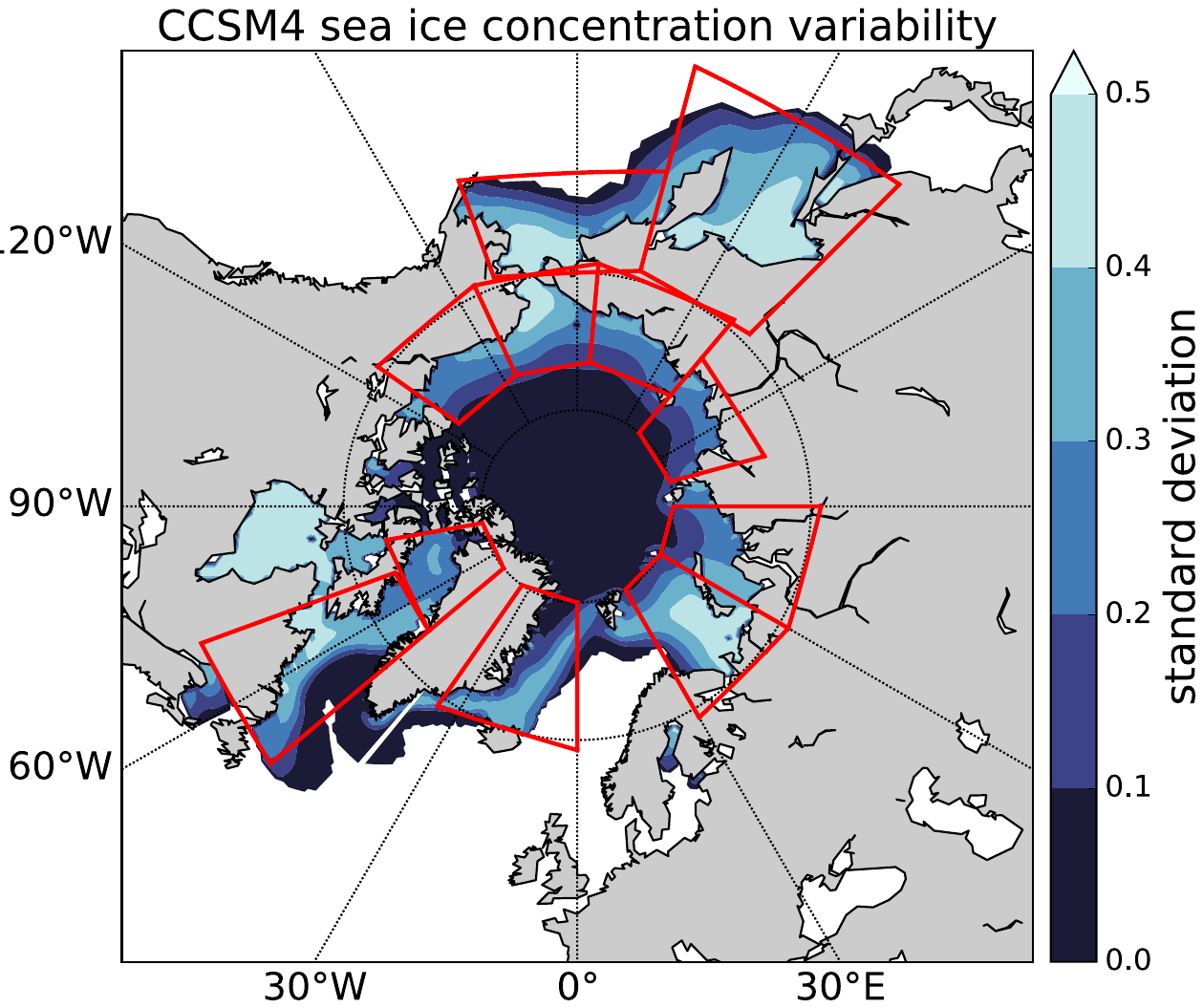}
\caption{Standard deviation of monthly sea ice concentration (SIC) from the CCSM4 control run, with regions considered in our forecasting: pan-Arctic (45N--90N), Beaufort Sea (155W--125W, 65N--75N), Chukchi Sea (175E--155W, 65N--75N), East Siberian Sea (140E--175E, 65N--75N), Laptev Sea (105E--140E, 70N--80N), Kara Sea (60E--90E, 65N--80N), Barents Sea (30E--60E, 65N--80N), Greenland Sea (35W--0E, 65N--80N), Baffin Bay (80W--50W, 70N--80N), Labrador Sea (70W--50W, 50--70N), Bering Sea (165E--160W, 55N--65N), and Sea of Okhotsk (135E--165E, 45N--65N).}
\label{fig:regions}
\end{figure}

\begin{figure*}
\centering
\noindent\includegraphics[width=174mm]{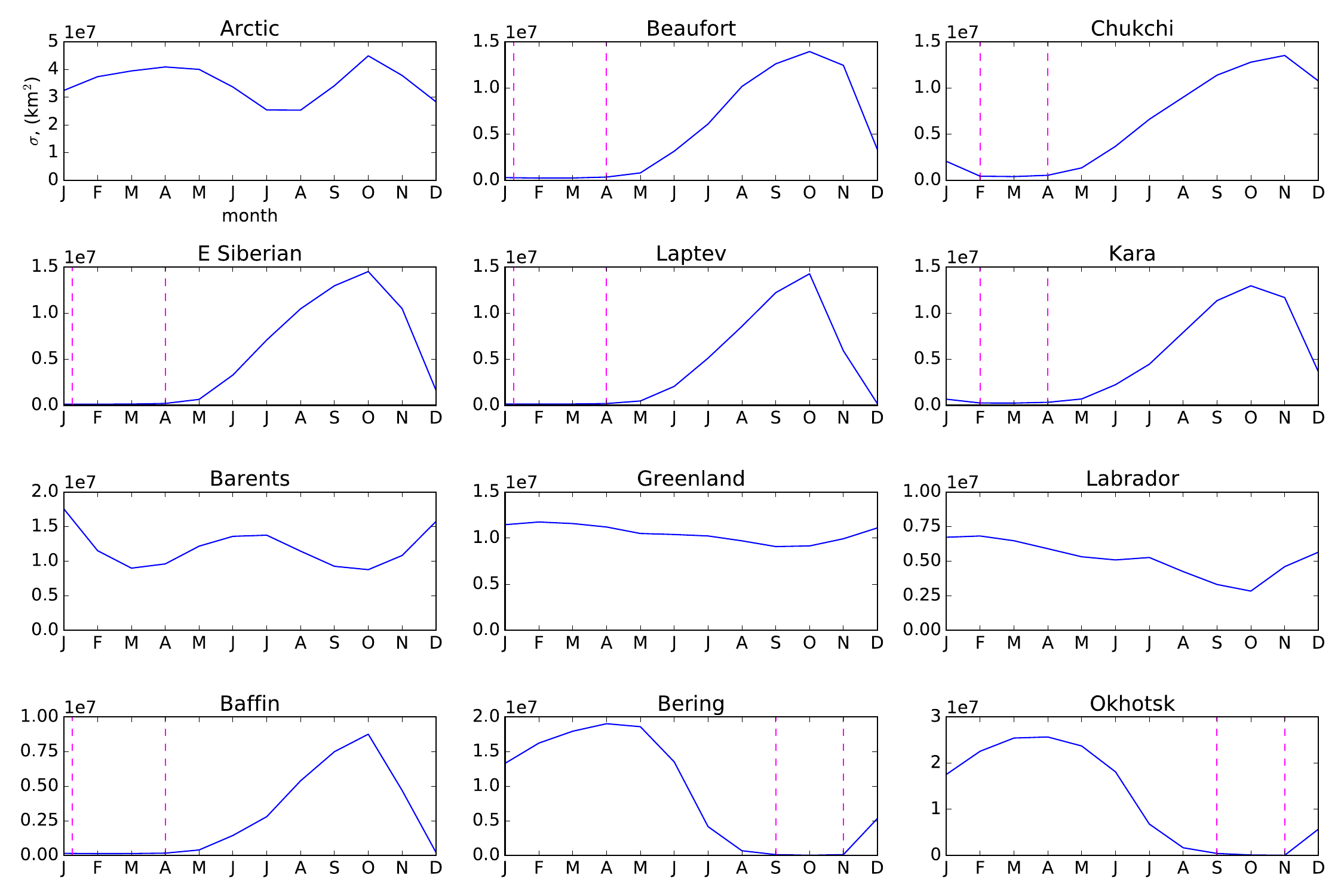}
\caption{Monthly standard deviation of sea ice area anomalies for the pan-Arctic and regions indicated in Fig.~\ref{fig:regions}. Periods of very low interannual variability, corresponding to months when the region is either completely ice covered (Beaufort, Chukchi, East Siberian, Laptev, Kara Seas, and Baffin Bay in late winter) or completely free of ice (Bering and Okhotsk seas in early fall), are indicated with magenta lines. Since the sea ice state is driven by climatology at these times, we do not expect skill in predicting anomalies from this state.}
\label{fig:region_std}
\end{figure*}

 We use monthly averaged CCSM4 \citep{gent2011community} model data from a pre-industrial control simulation (b40.1980), where 800 years of the simulation are split into a training dataset and a test dataset, 400 years each. The sea ice component is CICE4 \citep{hunke2008cice}, the ocean component is POP \citep{smith2010parallel}, and the atmosphere component is CAM4 \citep{neale2010description}. Our default experimental setup is to include SIC, SST, and SLP fields, and we will later explore the role of SIT as an additional predictor variable. We consider the entire Arctic, as well as the following regions: Beaufort Sea, Chukchi Sea, East Siberian Sea, Laptev Sea, Kara Sea, Barents Sea, Greenland Sea, Baffin Bay, Labrador Sea, Bering Sea, and Sea of Okhotsk. The regions are depicted in Fig.~\ref{fig:regions}, shown with this dataset's sea ice concentration variability calculated over the entire control run. Each region's monthly standard deviation in sea ice area anomalies are shown in Fig.~\ref{fig:region_std}. For regions and seasons with very low interannual variability, such as some central Arctic basins that are 100\% ice covered in late winter, or seasonal ice zones that are ice-free in the summer, we do not expect skill in predicting anomalies from this state. Note that despite that a pre-industrial control simulation is not fully indicative of our current transient climate, our objective is to establish a baseline of performance for KAF in predicting sea ice anomalies by making use of a large training data set of a climate without a secular trend, so that useful historical analogs may be identified and predictive skill robustly assessed. The Arctic sea ice anomalies from this dataset exhibit interannual variability, but no drift.

\par Our target observable $f$ for prediction is integrated anomalies in sea ice area and volume. Sea ice anomalies in the test data period are calculated relative to the monthly climatology calculated from the training data set. While this should not be a concern in a pre-industrial control run with no secular trend, it may be of more importance in other scenarios. Damped persistence forecasts are initialized with the true anomaly (as opposed to the out-of-sample extension value), so all forecasts will have initial error metrics greater than damped persistence due to reconstruction error.

\par We have considered various combinations of SIC with SST, SIT, and SLP as predictor variables, although most of the results presented here use the combination SIC, SST, \& SLP unless specified otherwise. The ice and ocean state variables are restricted to each region, whereas pan-Arctic SLP data is used for regional analysis to allow for possible teleconnection effects. While adding more variables, and thereby increasing the domain size and including more physics, should not result in the reduction of skill, in practice it may result in a loss of discriminating power of the kernel. A balance needs to be considered between the inclusion of variables that add more physics to the training data, and the ability of KAF to leverage this information in discerning useful historical analogs.

\par Regional predictions use training data only from that region, which does not account for predictive information outside the region boundaries that may advect across region boundaries. However, this approach does allow for better selection of historical analogs in that only local information is used in weighting analogs. In separate calculations, we have tested using pan-Arctic training data for predicting regional sea ice anomalies, and find better predictive skill when only regional data is used for training (with the exception of SLP).

\par An embedding window of $q=12$ months is used in constructing the kernels in Eq.~\eqref{eq:rowsum}; 6 and 24 month embedding windows were also tested for robustness, and while results were similar for a 6 month window, results with 24 months were marginally worse than 12 months. We use an ensemble size of 100 (number of non-zeros entries per row retained in $P$), which represents about 2\% of the total sample size, but the results are not sensitive to ensemble size (see \cite{comeau2017data}). Lastly, we use the shape parameter $\sigma_0=2$.

\section{Results}
\label{sec:results}
\subsection{Pan-Arctic} 
\begin{figure}
\centering
\noindent\includegraphics[width=84mm]{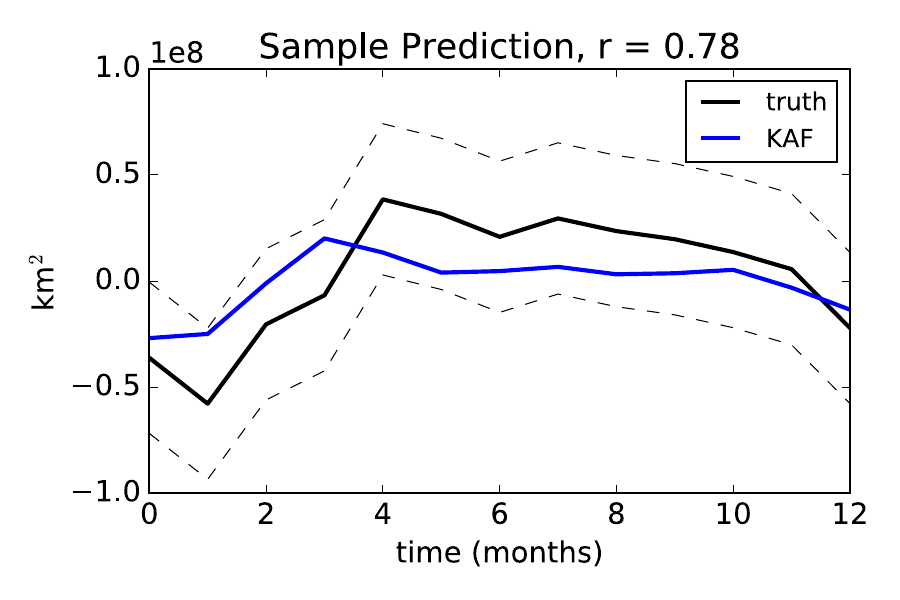}
\caption{Sample forecast trajectory of Arctic sea ice area anomalies using SIC, SST, and SLP as predictors. Dashed lines represent 1 standard deviation of the sea ice area anomaly over the test period.}
\label{fig:arctic_traj}
\end{figure}

\begin{figure}
\centering
\noindent\includegraphics[width=84mm]{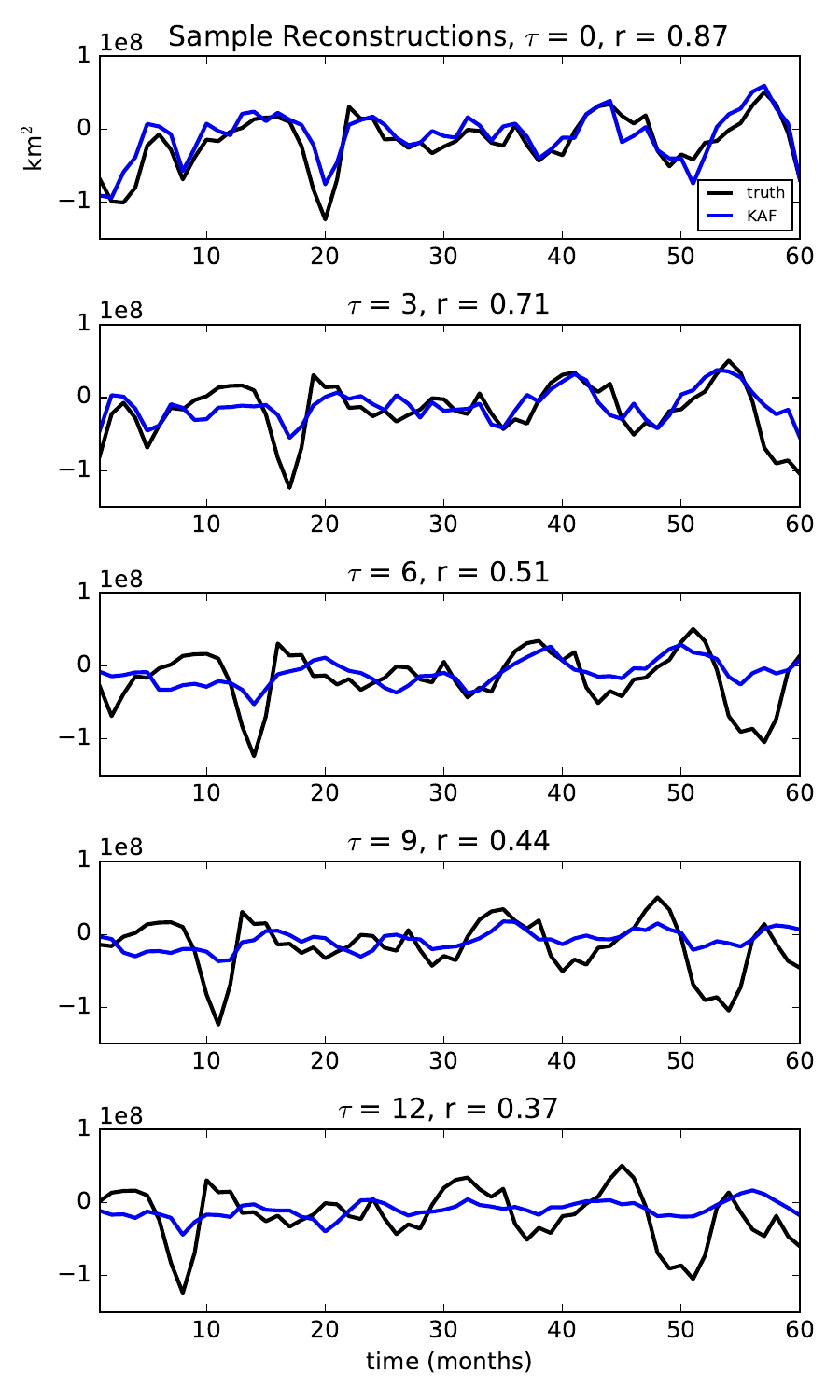}
\caption{Reconstructions at different time lags for Arctic sea ice area anomalies, taken over a random sample window from the forecast period, initialized in January. The blue forecast at each point is made from a lead time indicated by the panel. Degradation of forecast fidelity is seen as the lead time increases, though the particular results are dependent on the random sample window chosen.}
\label{fig:arctic_lags}
\end{figure}

We first focus on pan-Arctic sea ice area anomalies, using SIC, SST, and SLP predictor data, and a 12 month embedding window. Fig.~\ref{fig:arctic_traj} shows a sample forecast trajectory compared to the ground truth. While too much predictive value should not be inferred from single sample trajectories, it is common for forecasts to falter when near zero, as there is difficulty in determining the sign of the future anomaly when the state is very near climatology, even with dynamic information encoded into the forecasting scheme. We show the degradation of the forecasts as lead times increase in Fig.~\ref{fig:arctic_lags}, where forecasts are performed with 0, 3, 6, 9, and 12 month lead times. The initial reconstruction matches the truth reasonably well, and forecasts become increasingly smoothed out towards climatology with increasing lead time.

\par
\begin{figure}
\centering
\noindent\includegraphics[width=84mm]{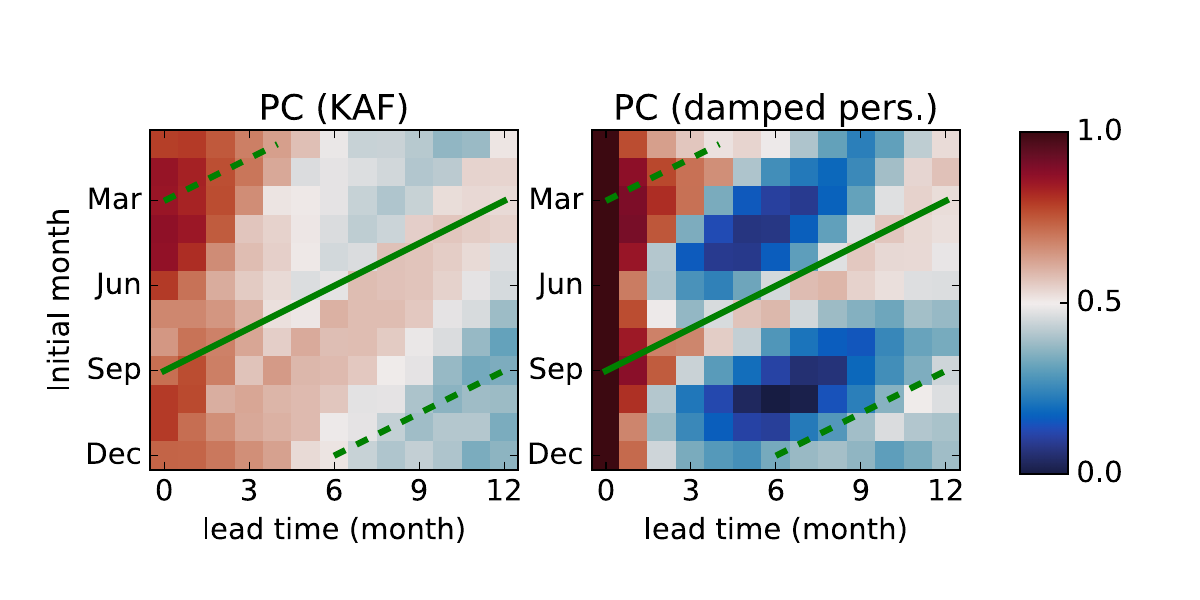}
\caption{Pattern correlation values as a function of lead time and initial month, for predictions of pan-Arctic sea ice area anomaly forecasts using KAF (left) with SIC, SST, and SLP as predictors, and damped persistence (right). Considering a pattern correlation of 0.5 as a threshold for predictive skill, red indicates predictive skill, and blue indicates lack of skill. The green solid line represents the expected 'melt-to-growth' sea ice reemergence limb. The green dashed line represents the expected 'growth-to-melt' sea ice reemergence limb.}
\label{fig:arctic_pc}
\end{figure}

 To quantify forecast skill, we consider the error metrics from Sect.~\ref{subsec:error_metrics} averaged over all forecasts initialized in the test period (400 years of monthly data, minus the length of the embedding window). In Fig.~\ref{fig:arctic_pc}, we show pattern correlation conditioned on initial month of prediction and lead time, for KAF and damped persistence forecasts as a benchmark for comparison. Lines corresponding to sea ice reemergence phenomena are overlaid in Fig.~\ref{fig:arctic_pc}. One line originates in September, and follows predictions symmetric about that month, meaning a prediction initialized $n$ months before September is targeting $n$ months after September. This represents the 'melt-to-growth' sea ice reemergence limb. A similar line is drawn originating from March, corresponding to the 'growth-to-melt' sea ice reemergence limb. Increased skill is expected to appear along these lines due to sea ice reemergence aiding the predictions.

\par These reemergence lines align more closely with the damped persistence forecasts areas of success than with KAF, and in particular the March limb does not correspond to predictive skill beyond 6 months in the KAF forecasts. However, a wider band of skill in the KAF forecasts envelopes the 'melt-to-growth' limb than in the damped persistence forecast. Note that due to the reconstruction error at lead time 0 in KAF forecasts, we would not necessarily expect exact alignment with the reemergence limbs.

\par Beyond initial reconstruction, KAF generally outperforms damped persistence and is above the 0.5 threshold for almost all of the first 6 months predicted range, including out to 12 months along the 'melt-to-growth' (solid line) reemergence limb.
Damped persistence generally loses skill after 1--2 months, with a couple of exceptions which remain skillful along reemergence limbs. The largest differences between the two forecasting methods appears between the reemergence limbs, where KAF has notably higher pattern correlation values. Damped persistence appears to be strongly impacted by the summer predictability barrier, as predictions from summer have skill for only very short lead times.

\subsection{Regional Arctic}

\begin{figure*}
\centering
\noindent\includegraphics[width=174mm]{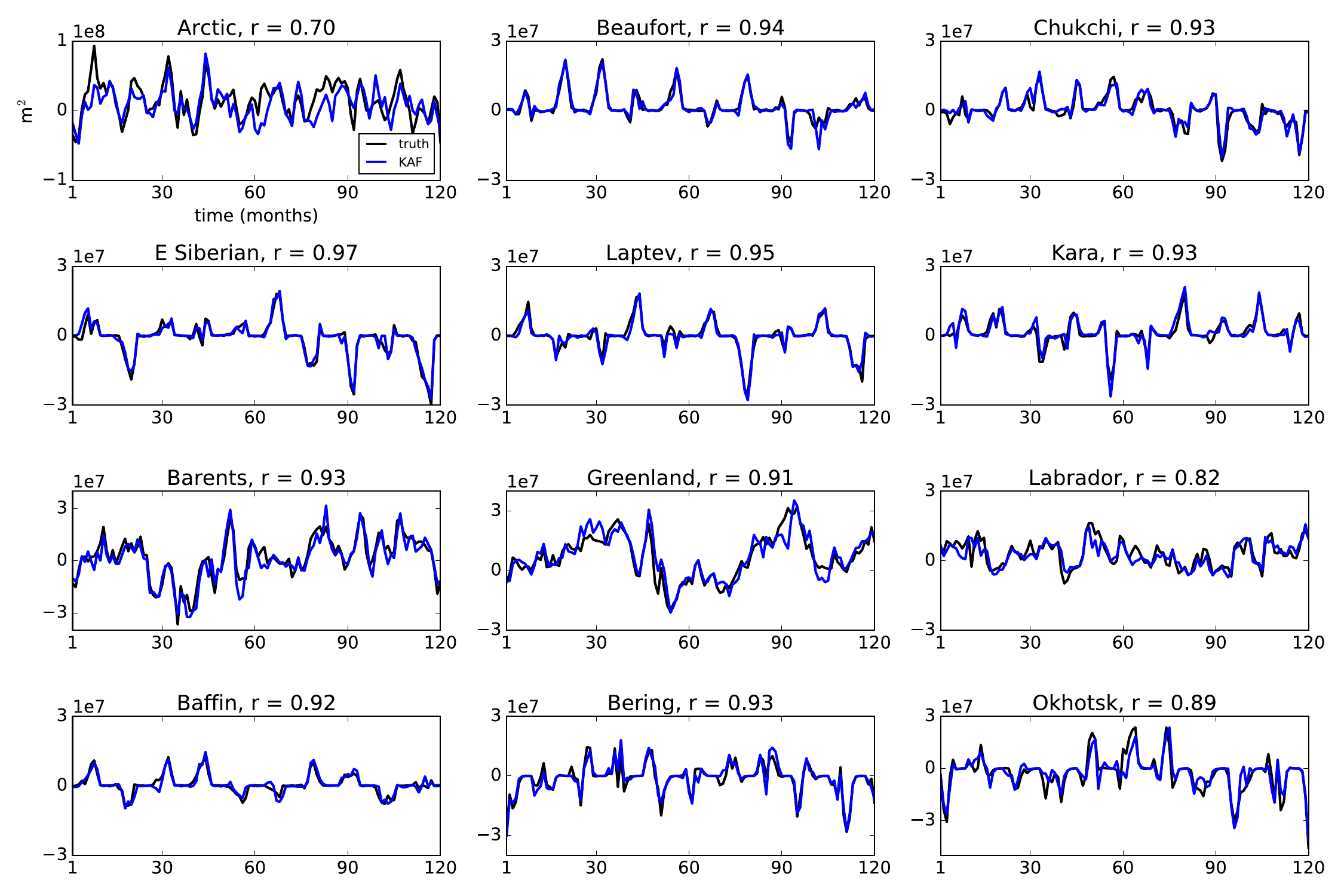}
\caption{Initial reconstructions (lead time 0) of regional sea ice area anomalies compared to truth, taken as a random sample from the forecast period, initialized in January. The regions that have more persistent anomalies (Barents and Greenland), are North Atlantic adjacent, which has been found in other studies to be regions of relatively high predictability.}
\label{fig:regional_truth}
\end{figure*}

\par While predicting pan-Arctic sea ice area minimums and maximums has been of great interest, as more areas of the Arctic become accessible, an increased effort has been made in regional scale predictions. Snapshots of regional sea ice anomalies (calculated against regional climatologies) in Fig.~\ref{fig:regional_truth} demonstrate different behavior around the Arctic basin. The out-of-sample extension values are plotted with the truth, and again should be thought of as the lead time 0 forecast. The central Arctic basins (Beaufort, Chukchi, East Siberian, Laptev, Kara Seas, and Baffin Bay) experience winter months with near zero anomaly as they are 100\% ice covered. Continuing westward to the Barents Sea, we begin to see the strong influence of the North Atlantic in regulating sea ice cover. More persistent anomalies are seen in the Barents and Greenland seas, which we see later leads to greater predictability (Figs. \ref{fig:region_pc} and \ref{fig:pc_TM}). Moving across to the North Pacific basins, the Bering Sea and Sea of Okhotsk also exhibit regular intervals of near zero anomaly due to being completely ice free in the summer months.

\begin{figure*}
\centering
\noindent\includegraphics[width=174mm]{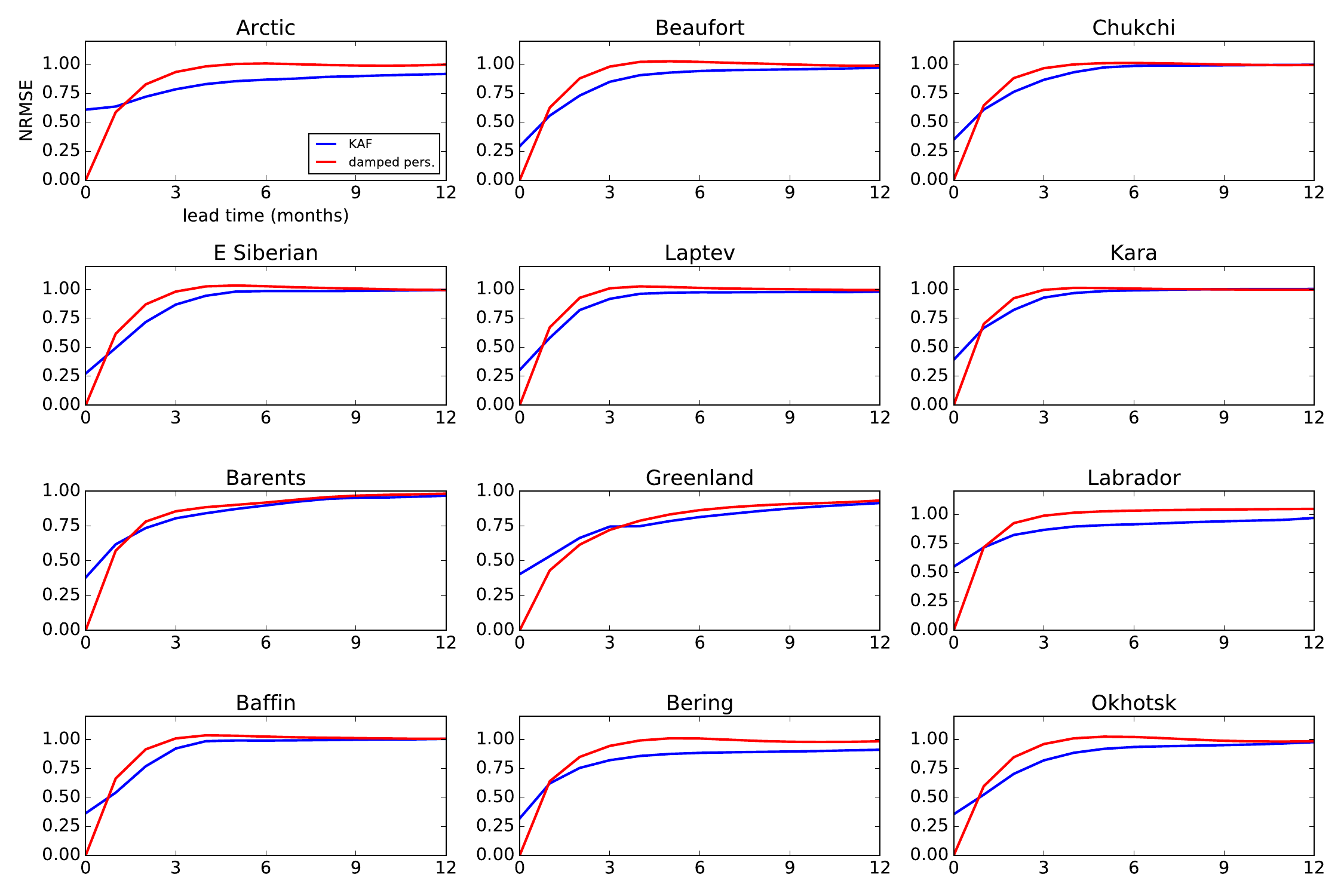}
\caption{Normalized RMSE for regional sea ice area anomaly predictions, averaged over all months, for KAF and damped persistence as a benchmark. As NRMSE scores approach one, predictive skill is considered to be lost. KAF suffers reconstruction errors at lead time 0, then outperforms damped persistence usually after one month, followed by a loss of predictive skill around 3 or 4 months for most regions. The loss of predictive skill in the pan-Arctic forecast is notably slower than the regional counterparts.}
\label{fig:region_rms}
\end{figure*}

\begin{figure*}
\centering
\noindent\includegraphics[width=174mm]{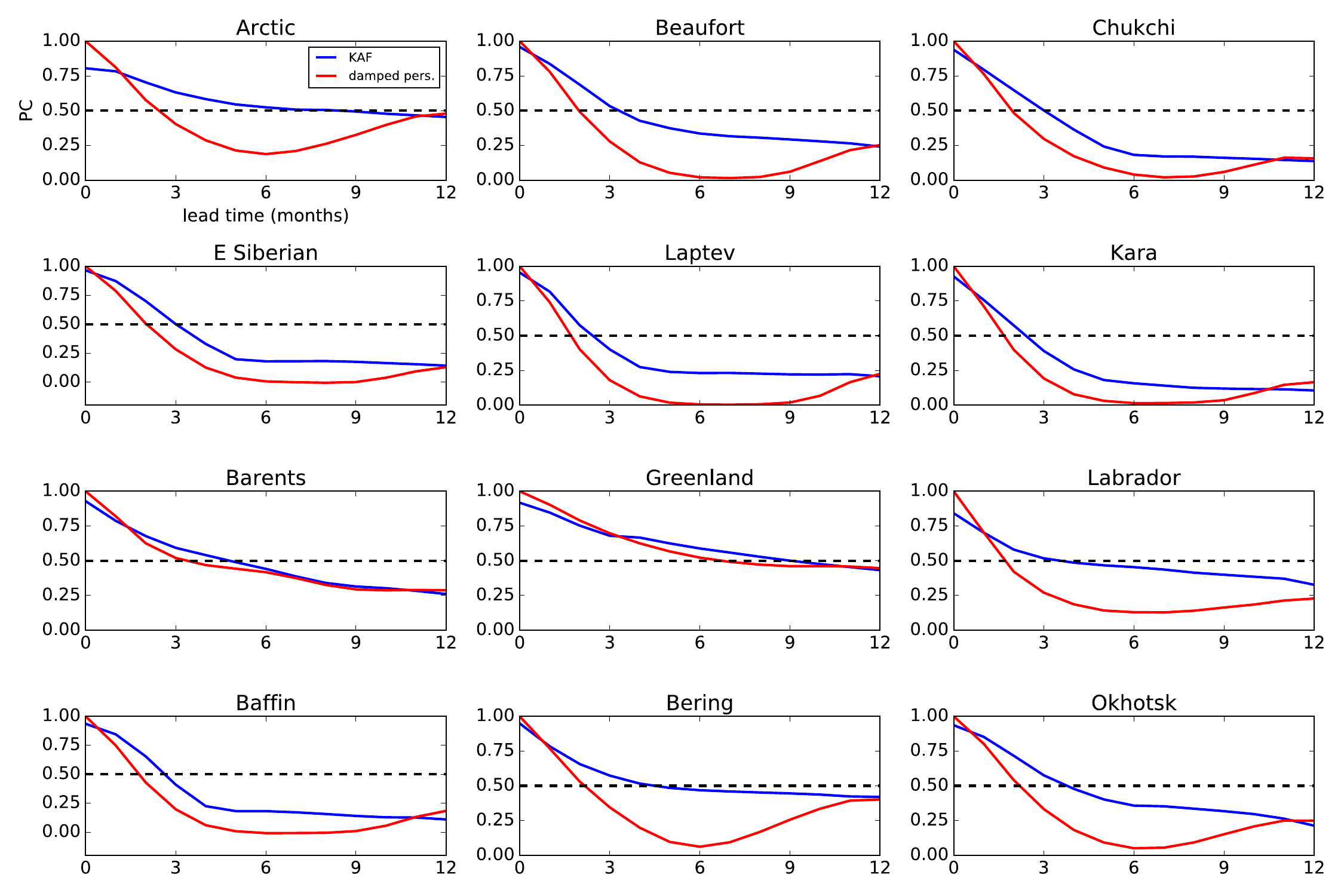}
\caption{Pattern correlation scores for regional sea ice area anomaly prediction by region, averaged over all months. Reconstruction errors are less noticeable in this metric (apart from pan-Arctic), and KAF exceeds damped persistence in almost every region and lead time. Note the North Atlantic regions are the most persistent, as demonstrated by slow decay of pattern correlation. There, regional improvements over damped persistence are marginal, though in pan-Arctic, the improvement is several months.}
\label{fig:region_pc}
\end{figure*}

\par The aggregated error metrics, averaged over all months for each region in Figs.~\ref{fig:region_rms} (NRMSE) and  \ref{fig:region_pc} (pattern correlation) show that KAF consistently outperforms damped persistence (or at least fares no worse) once an initial reconstruction error is overcome, typically after one month. The KAF NRMSE approaches 1 around the same time its pattern correlation score drops below 0.5, two measures of predictive skill being lost, which for most regions occurs around 3 or 4 months lead time. Disregarding pattern correlation scores below the 0.5 threshold may cut into some apparent gains of KAF over damped persistence, but it is worth noting the decay rate of KAF pattern correlation is slower than damped persistence, sometimes dramatically so (e.g. Bering and Labrador). The persistent nature of the North Atlantic adjacent basins seen in Fig.~\ref{fig:regional_truth} manifests itself as slower than average decay of damped persistence. 

\begin{figure*}
\centering
\noindent\includegraphics[width=174mm]{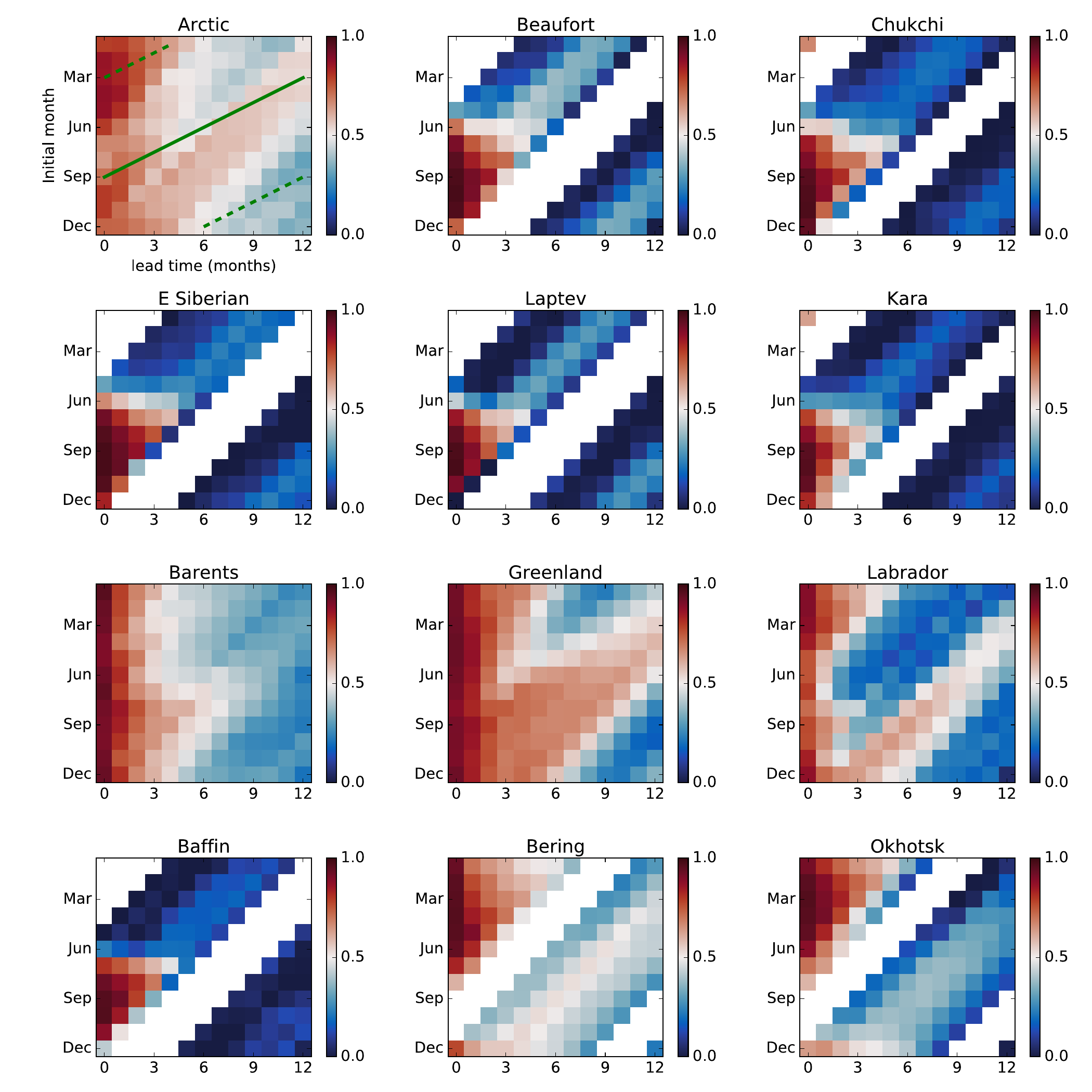}
\caption{Pattern correlation for KAF forecasts of sea ice area anomalies as a function of initial month and lead time by region. Pan-Arctic values are reproduced from the left panel of Fig.~\ref{fig:arctic_pc}. Months when the region has near zero interannual variability, as seen in Fig.~\ref{fig:region_std}, have been whited out. Green lines correspond to sea ice reemergence limbs, as in Fig.~\ref{fig:arctic_pc}.}
\label{fig:pc_TM}
\end{figure*}

\par Conditioning forecasts on the initial month of prediction allows us to parse out seasonal impacts on predictability. The combined spatial and temporal effects of predictability highlight particularly skillful months and regions to predict, as seen in Fig.~\ref{fig:pc_TM}. The regions and seasons of near zero interannual variability identified in Fig.~\ref{fig:region_std} are whited out, as no skill is expected in predicting anomalies.

\par Starting with the central Arctic basins (Beaufort, Chukchi, E. Siberian, Laptev, and Kara Seas), we see that if predictions are skillful, it is  only for short lead times - up to 3 months at most. These predictions clearly are impacted by the summer predictability barrier, as there is very poor skill for predictions initialized before July. For these regions, sea ice reemergence along the March limb is providing some increase in skill. This is seen along horizontal rows where skill increases from dark blue to light blue, however this increase is not enough for the forecasts to be considered skillful.

\par Moving to the North Atlantic adjacent basins (Barents, Greenland \& Labrador Seas), we see significant skill at longer lead times, perhaps the manifestation of longer persistence in anomalies seen in Fig.~\ref{fig:regional_truth}. In the North Pacific basins (Bering Sea and Sea of Okhotsk), an increase in skill can be seen along the later months of the September reemergence limb (6-12 months), resulting in pattern correlation skill scores close to or exceeding our 0.5 threshold. For this set of central Arctic and North Pacific basins, note that the periods of highest interannual variability (Fig.~\ref{fig:region_std}) correspond to periods when KAF exhibits the highest skill.

\par
\begin{figure*}
\centering
\noindent\includegraphics[width=174mm]{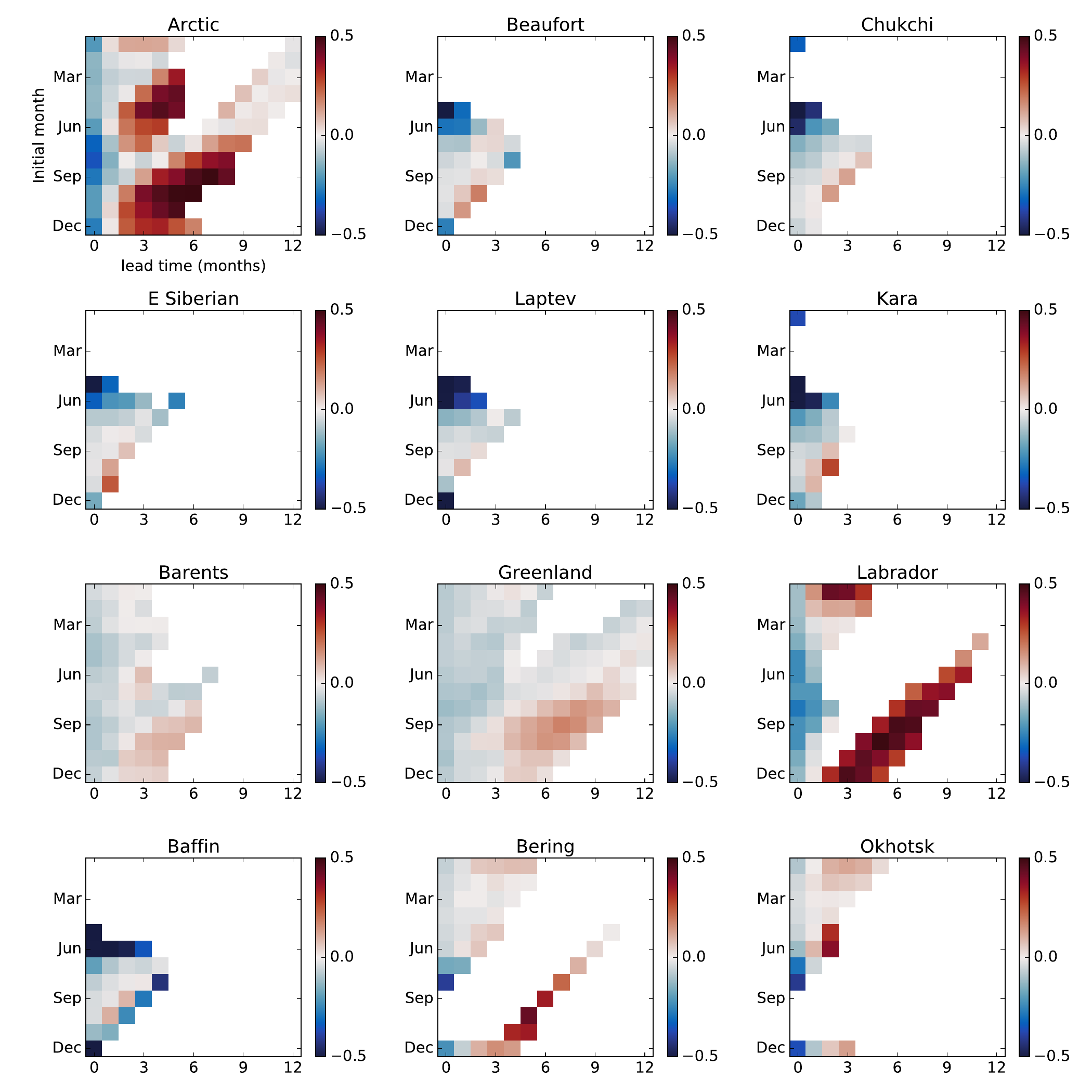}
\caption{Difference in pattern correlation scores of KAF over damped persistence to illustrate the gain in predictive skill, with zero in place of any value where both scores are below 0.5. Red indicates KAF outperforming damped persistence, and blue vice-versa.}
\label{fig:pc_TMdiff}
\end{figure*}

 To demonstrate the gain in predictive skill of KAF over damped persistence, rather than plot damped persistence pattern correlation by initial month as in Fig.~\ref{fig:arctic_pc}, we instead plot the difference in pattern correlation scores, KAF minus damped persistence (Fig.~\ref{fig:pc_TMdiff}). We zero out any value where both pattern correlation scores are below the threshold of 0.5, which we consider as not indicative of predictive skill. Considerable improvement over damped persistence is seen in pan-Arctic forecasts with lead times of 3--6 months, as well as in some marginal ice zones for predicting late winter, most notably in the Labrador sea. The skill KAF has in the central Arctic basins for fall months is mostly matched or exceeded by damped persistence.

\subsection{Role of predictor variables}

\par
\begin{figure}
\centering
\noindent\includegraphics[width=84mm]{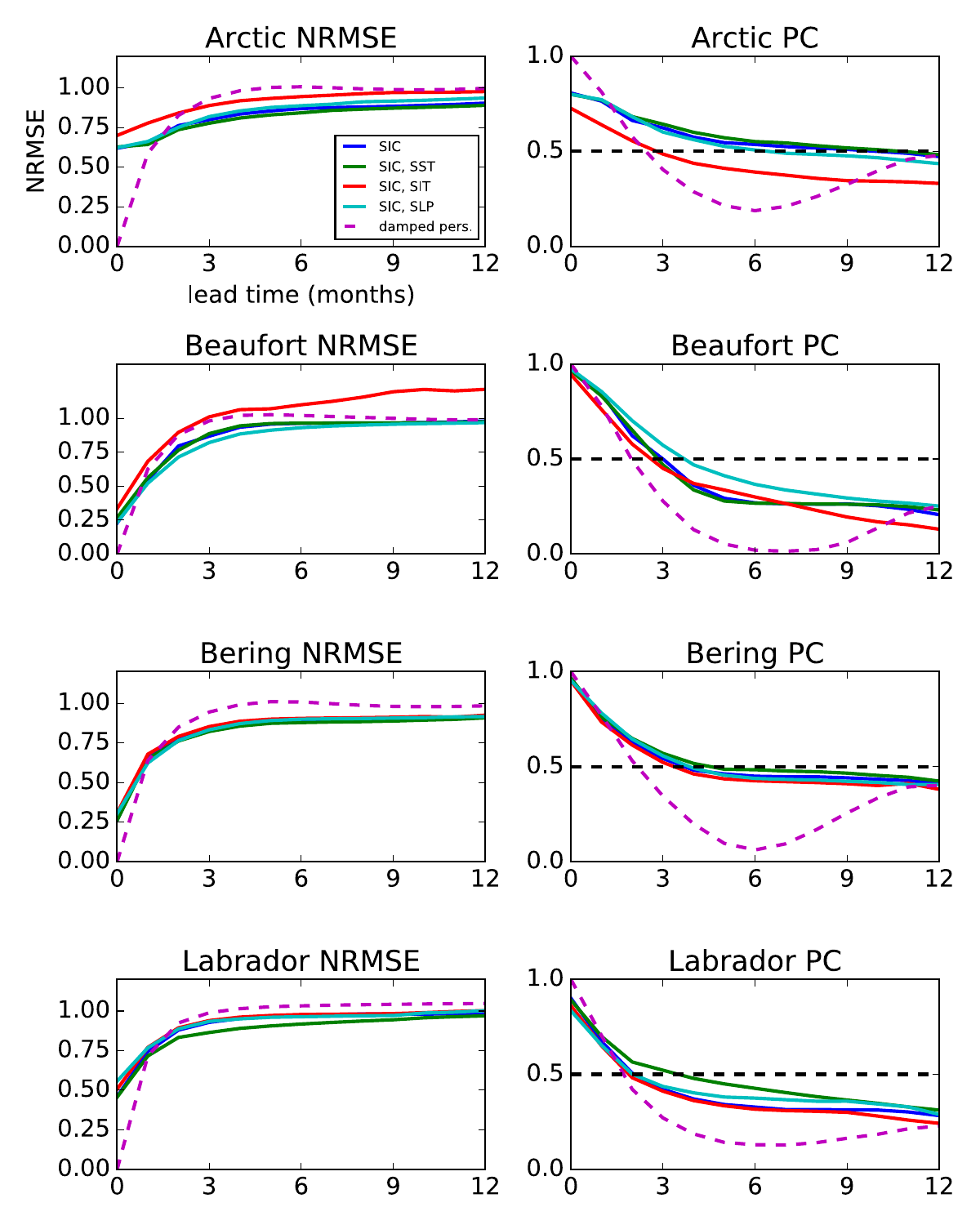}
\caption{Prediction results for Arctic sea ice area anomalies using different predictor variables for the pan-Arctic, a region in a mainly perennial ice zone (Beaufort Sea), a region in a marginal ice zone (Bering Sea), and a North Atlantic region with strong memory from persistence (Labrador Sea). Most of the skill is from SIC alone, although SLP aids the Beaufort sea forecasts, while SST aids the Labrador sea forecasts.}
\label{fig:arctic_varsUsed}
\end{figure}

 So far, the experiments we have shown have used SIC, SST, and (pan-Arctic) SLP as predictor variables, from which kernel evaluations to determine similarity are based (in Takens embedding space). To address the predictive power of each of these variables, in Fig.~\ref{fig:arctic_varsUsed} we show the effect of combinations of SIC with each of SST, SLP, and SIC separately as predictors for the pan-Arctic, as well as a representative perennial ice zone (Beaufort), marginal ice zone (Bering) ice zone, and a North Atlantic adjacent basin (Labrador). In general, we find that KAF extracts much of its predictive power through SIC alone, with modest gains, or at times losses, when including an additional predictor. For example, including SLP as a predictor variable increases the Beaufort sea forecasts by about a month over those using SIC alone, and similarly for SST in the Labrador sea forecasts. Interestingly, adding sea ice thickness information can actually be detrimental to sea ice area anomaly prediction, as seen in the pan-Arctic forecasts. This may seem surprising, given other studies' emphasis on the importance of sea ice thickness measurements. However, in the context of kernel evaluation, increasing the dimension of our state vector may yield less discernible informative historical analogs.  A similar degradation of performance when including SIT data in the kernel was observed in the study of \citet{bushuk2017seasonality} on SIT-SIC reemergence mechanisms. This behavior was attributed to the slower characteristic timescale of SIT data, resulting in this variable dominating the phase velocity-dependent kernel in Eq.~\eqref{eq:nlsakernel_multivariate}. 

\par
\begin{figure}
\centering
\noindent\includegraphics[width=84mm]{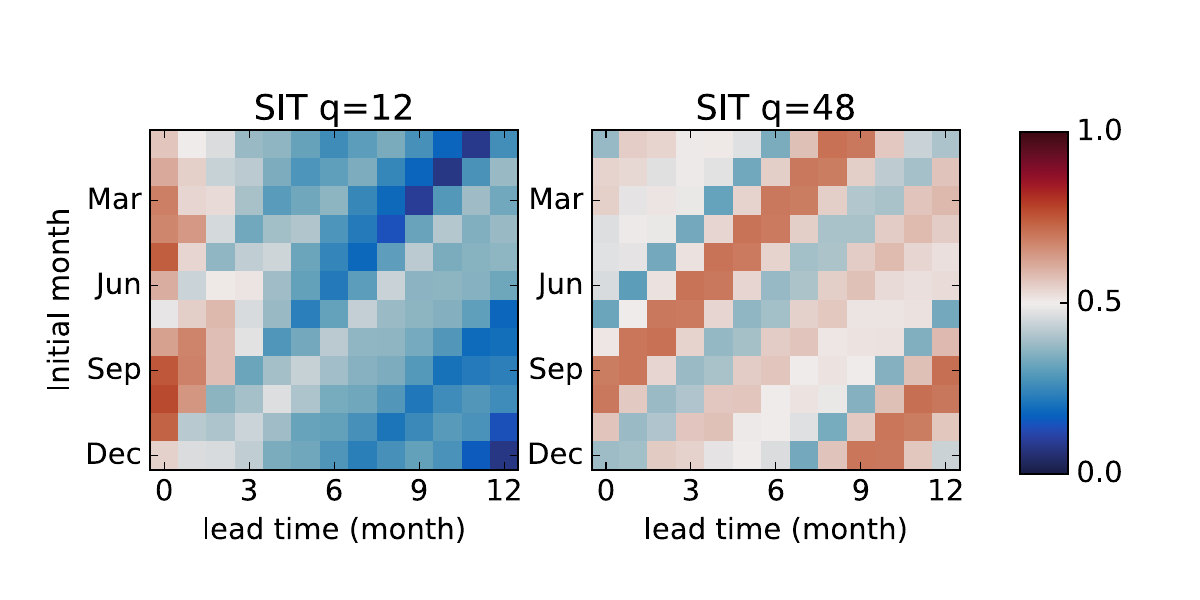}
\caption{Prediction results (pattern correlation scores) for pan-Arctic sea ice area anomalies using SIC, SST, and SIT (instead of our default SIC, SST, and SLP) predictors with different embedding windows for SIT; $q=12$ (left) and $q=48$ (right) months.}
\label{fig:arctic_SITq48}
\end{figure}

 This degradation of skill could in part be mitigated by allowing for a longer embedding window for SIT. Fig.~\ref{fig:arctic_SITq48} shows pattern correlation scores for sea ice area anomalies using SIC, SST, and SIT predictors using two different embedding windows for SIT: $q=12$ and $q=48$. The $q=12$ case in general shows less skill than our default experimental setup using SLP in place of SIT (See Fig.~\ref{fig:arctic_pc}). For the $q=48$ case, an interesting pattern appears where prediction skill is fairly constant for predicting a particular month, which is seen by following lines of slope 1. The pattern of skill in predicting a particular month closely follows the variance of Arctic sea ice area by month (see Fig.2), with months of higher variance corresponding to higher skill. With an embedding window for SIT set to 4 times the length of the prediction horizon, the decay in predictive skill is effectively not seen within the given 12 month prediction horizon.

\par Turning to an atmospheric predictor variable, while the inclusion of pan-Arctic SLP does not hamper our prediction skill, it offers only marginal improvement. An exception to this is the Beaufort Sea, which experiences a gain of one month in predictive skill with the inclusion of SLP (Fig.~\ref{fig:arctic_varsUsed}). This marginal predictive power of SLP is most likely due to the fact that the quantities used are monthly averaged, and perhaps too temporally coarse to reflect the chaotic atmospheric influence on sea ice cover on shorter time scales, or that SLP itself is not predictable on month long time scales.

\subsection{Regional volume anomalies}

\begin{figure*}
\centering
\noindent\includegraphics[width=174mm]{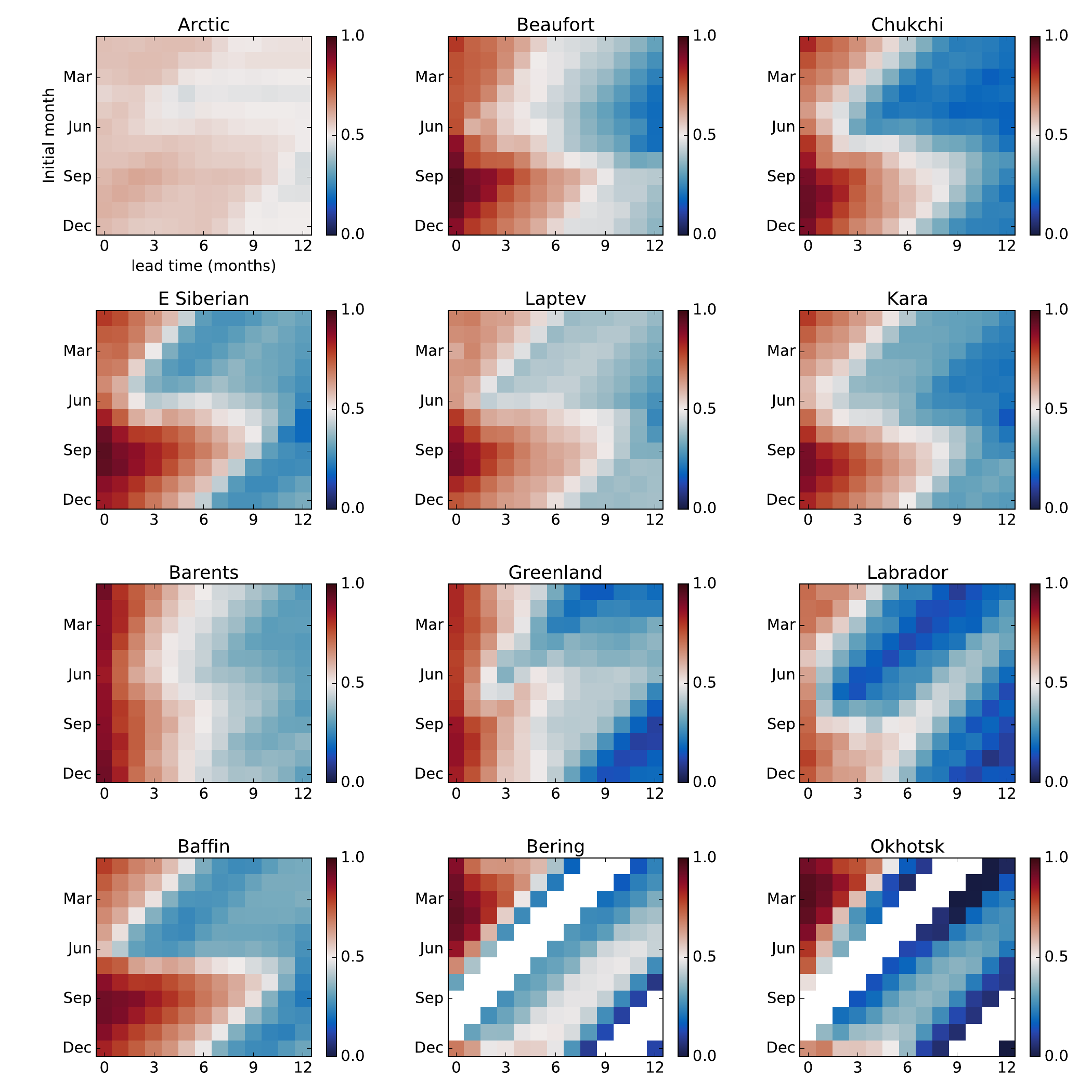}
\caption{Forecasts for Arctic sea ice volume anomalies, predicted using only SIC, SST, and SLP data. By not using SIT as a predictor variable, we are predicting an unobserved quantity, and damped persistence forecasts based on the true anomaly outperform KAF on all spatial and temporal scales (not shown). Times of near zero interannual variability corresponding to ice free regions and months are whited out.}
\label{fig:iva_noSIT_pc_TM}
\end{figure*}

\par We also consider the problem of forecasting sea ice volume anomalies, which in general show more persistence than sea ice area anomalies. The reason in part is due to thinner ice being more sensitive to advection by winds to areas that are more or less prone to melting, and this thin ice drives area anomalies. Fig.~\ref{fig:iva_noSIT_pc_TM} shows regional forecast pattern correlation scores for predicting sea ice volume anomalies, having only observed SIC, SST, and SLP, following the same implementation as for area forecasts. In this example we are predicting an unobserved variable, yet see skill for lead times as high as 9 months in some regions. However, these do not compare favorably against a damped persistence forecast using the ground truth (not shown) due to inherent persistence of volume anomalies, though this would also not be a fair comparison given the KAF forecasts are not observing the full observable. The summer melt predictability barrier is clearly seen here as a sharp decline in skill from June to July.

\par
\begin{figure*}
\centering
\noindent\includegraphics[width=174mm]{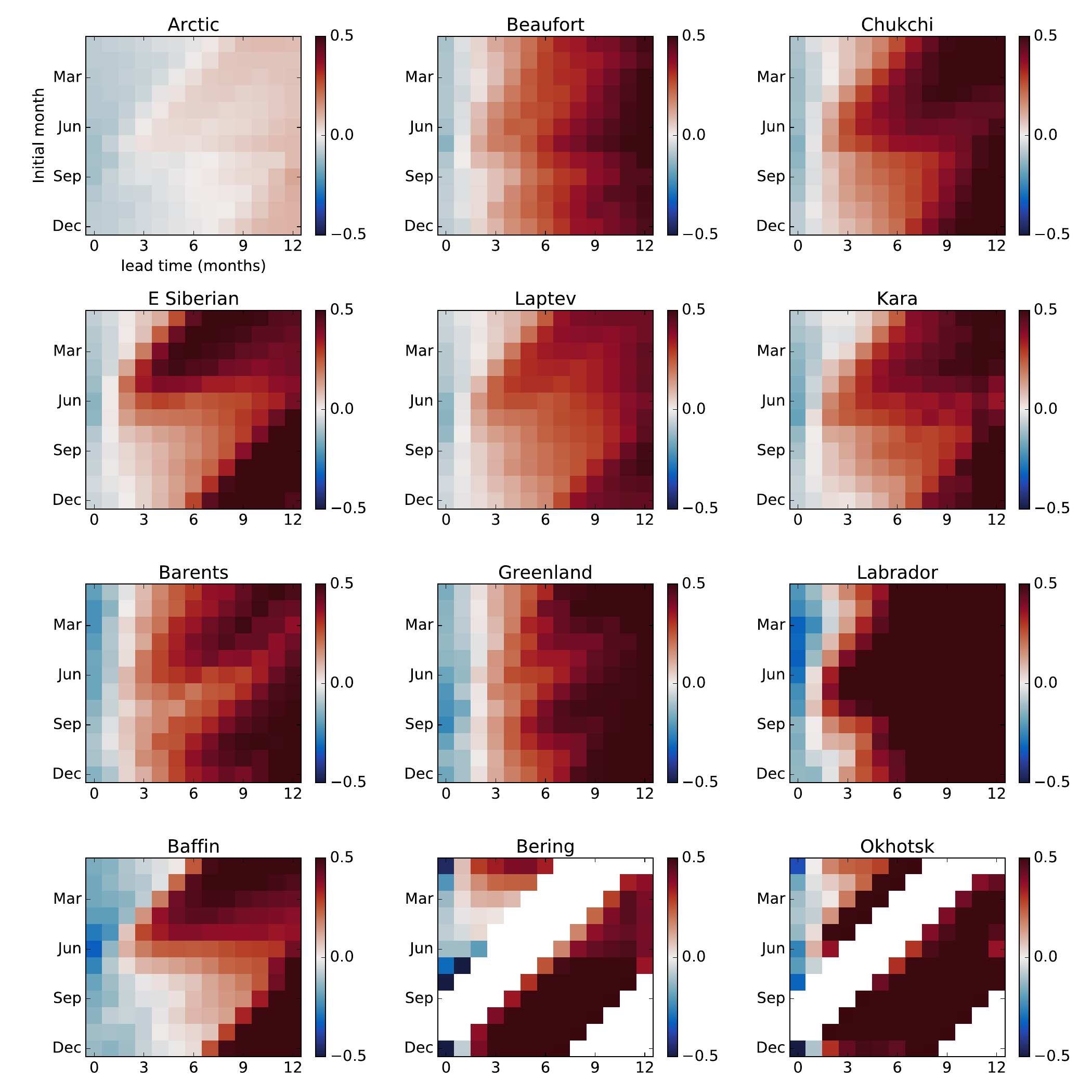}
\caption{Difference in pattern correlation score of KAF over damped persistence for predicting sea ice volume anomalies, with predictor variables SIC, SST, and SIT. SIC and SST have the default $q=12$ month embedding window, whereas SIT has an extended embedding window of $q=48$ months. Whited out areas indicate that both KAF and damped persistence pattern correlation is below 0.5, as well as areas of near zero interannual variability.}
\label{fig:iva_SIT_pc_TM}
\end{figure*}

 When we include SIT as predictor data with an increased embedding window of $q=48$ months, we expectedly see a substantial increase in skill in predicting sea ice volume anomalies. In Fig.~\ref{fig:iva_SIT_pc_TM}, we show the difference in pattern correlation score of KAF over damped persistence (similar to Fig.~\ref{fig:pc_TMdiff}, but for volume anomalies). Damped persistence outperforms KAF at short lead times (0--2 months), largely due to the initial reconstruction error in KAF. For longer lead time (3--12 months), KAF retains predictive skill with pattern correlation scores that far exceed those of damped persistence in many regional forecasts. This gain in predictive skill in regional forecasts does not translate to pan-Arctic forecasts, where the difference between damped persistence and KAF is quite small, but follows the pattern that damped persistence scores higher at short lead times (0--6 months), and KAF scores higher at longer lead times (6--12 months). Pan-Arctic sea ice volume anomalies are to a large extent thermodynamically driven, as opposed to regional volume anomalies which also have dynamic effects of sea ice advecting across region boundaries. Thus pan-Arctic sea ice volume has a much longer time-scale of persistence than regional sea ice volume anomalies, and this benefits the damped persistence forecast, which is controlled by the lag-1 autocorrelation coefficient of the anomaly time series.

\section{Discussion \& Conclusions}
\label{sec:conclusions}
\par In this paper, we utilized KAF \citep{zhao2016analog,comeau2017data,alexander2017kernel}, a nonparametric method using weighted ensembles of analogs, to predict Arctic sea ice area and volume anomalies in CCSM4, for both pan-Arctic and regional scales, examining the effects of including SIC, SST, SLP, and SIT as predictors for our method. We find in general that for predicting pan-Arctic sea ice area anomalies, KAF outperforms the damped persistence forecast, or at minimum does not perform worse (with the exception of the inherent lag 0 reconstruction error), and the outperformance lead times range between 1 and 9 months, depending on region and season. Moving to regional scale basins and conditioning on the initial month of prediction, we see clear regional-seasonal domains when KAF succeeds at shorter lead times (3--4 months), as well as those when it fails (along with damped persistence).

\par For longer lead times, we found that while sea ice reemergence aided in predictive skill, this aid was often not enough to allow us to consider forecasts skillful. This may be due to the nature of these sea ice reemergence phenomena being centered around months of complete or zero ice coverage. The lead times needed to span this season for anomalies to reemerge are then too long, after KAF has already lost predictive skill. This could be a general reason why sea ice reemergence is not particularly helpful in aiding forecast skill, at least with respect to time-averaged skill metrics. On the other hand, another factor to consider is that sea ice reemergence itself has an interannual character, and conditioning forecasts only on years with active sea ice reemergence may yield an increase in skill along these limbs.

\par The North Atlantic seems to have a strong impact on sea ice area anomalies, as the adjacent regions (Barents, Greenland, and to a lesser extent, Labrador Seas) exhibit the strongest persistent anomalies, and have the highest year-round predictability. Predicting late winter/early spring in Greenland and Labrador seas in particular are examples of KAF success, which are skillful out to 12 months lead time. In general, KAF seems to do well at periods of high variability, despite the inherent penalty associated with a reconstruction error.

\par We find most of the predictive information for sea ice area is in SIC alone, with each of SST, SLP and SIT providing marginal improvements, although in some cases the inclusion of SIT actually hampers predictive skill. While we have success in reconstructing sea ice volume anomalies without using SIT as a predictor at the regional level, we see drastically improved performance with the inclusion of SIT in predicting pan-Arctic volume anomalies, particularly at the regional level, where forecasts remain skillful at 12 month lead times.

\begin{figure}
\centering
\noindent\includegraphics[width=84mm]{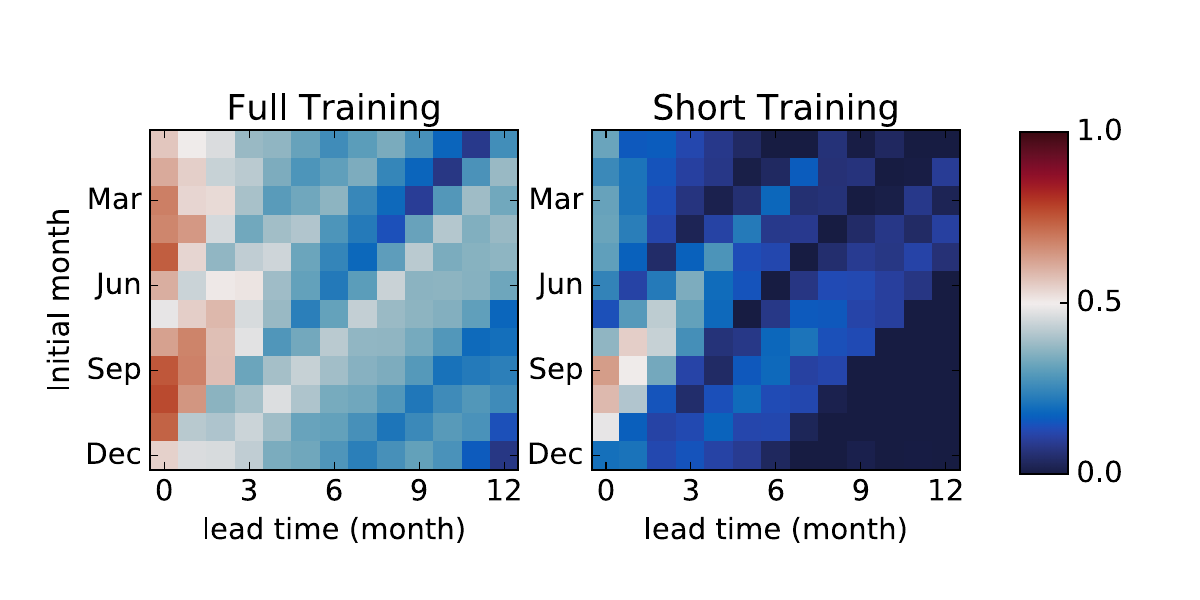}
\caption{Prediction results (pattern correlation scores) using SIC, SST, and SIT predictors for Arctic sea ice area anomalies comparing different length training records; full training of 400 years, and short training of only 40 years (similar length to the satellite record). Predictive skill is significantly hampered when the training record is not long enough to contain a suitable set of analogs.}
\label{fig:arctic_short_record}
\end{figure} 

\par Ultimately, the goal is to move to an operational prediction based on observational data, for which this is a first step. By using model data, we are able to make use of a long control run that has sampled the climate's natural variability and perform statistically robust estimates of skill. Limitations on the quality (i.e. presence of model biases or observation errors) and the length of training data will impact the performance of KAF, as it would any other statistical method. Experiments with much shorter lengths of control training data (e.g. 40 years) show a sharp decrease in KAF predictive skill (Fig.~\ref{fig:arctic_short_record}), underscoring the need for a rich enough set of training data where the system's full internal variability has been explored, even without the presence of a changing climate. Utilizing KAF to predict internal variability in conjunction with some method to account for the changing mean Arctic state would implicitly assume the internal climate variability itself is not changing, which also merits consideration. Another possible utility of KAF is in bias correction for a dynamical model forecast, similar to \citet{liu2017improving}. Furthermore, using multiple sources of information, such as multiple models or ensemble runs, may help mitigate biases from an individual model, and/or the need to collect training data over long time intervals.

\par Our future research plan is to use NLSA to extract an underlying `trend' in the data as a way of non-parametrically determining a trend (as opposed to fitting a linear or quadratic regression). This trend could then be extended to a forecast time using some form of extrapolation or out-of-sample extension technique, while the anomalies from this trend would be forecasted by the KAF method using datasets from control model runs as in this study. Other research directions include using a blended damped persistence and analog forecasting approach to avoid the initial reconstruction errors at short time scales, as well as forecasts using kernels targeted at specific observables.


\begin{acknowledgements}
The research of Andrew Majda and Dimitrios Giannakis is partially supported by ONR MURI grant 25-74200-F7112. Darin Comeau was supported as a postdoctoral fellow through this grant. Dimitrios Giannakis and Zhizhen Zhao are partially supported by NSF grant DMS-1521775. Dimitrios Giannakis also acknowledges support from ONR grant N00014-14-1-0150. Darin Comeau also acknowledges additional support from Regional and Global Climate Modeling program of the U.S. Department of Energy Office of Science, as a contribution to the HiLAT Project. We thank Mitch Bushuk for helpful discussions. We also thank two anonymous reviewers for their helpful comments in reviewing this manuscript.
\end{acknowledgements}



%
%
%

\end{document}